\documentclass[apjl]{emulateapj}
\usepackage{epsfig,bm,graphics,graphicx}

\advance \voffset by 0.00cm\relax

\submitted{Submitted to The Astrophysical Journal, Letters (Feb 9, 2004)}

\shorttitle{GRAVITATIONAL RADIATION RECOIL REVISITED}
\shortauthors{FAVATA, HUGHES, and HOLZ}

\begin{document}

\title{How black holes get their kicks: Gravitational radiation
recoil revisited}

\author{Marc Favata\altaffilmark{1}, Scott A.\ Hughes\altaffilmark{2},
and Daniel E.\ Holz\altaffilmark{3}}

\altaffiltext{1}{Department of Astronomy, Cornell University, Ithaca,
NY 14853}

\altaffiltext{2}{Department of Physics, Massachusetts Institute of
Technology, Cambridge, MA 02139}

\altaffiltext{3}{Center for Cosmological Physics, University of
Chicago, Chicago, IL 60637}

\begin{abstract}
Gravitational waves from the coalescence of binary black holes carry
away linear momentum, causing center of mass recoil.  This ``radiation
rocket'' effect has important implications for systems with escape
speeds of order the recoil velocity.  We revisit this
problem using black hole perturbation theory, treating the binary as a
test mass spiraling into a spinning hole.  For extreme mass ratios
($q\equiv m_1/m_2 \ll 1$) we compute the recoil for the slow inspiral
epoch of binary coalescence very accurately; these results can be
extrapolated to $q \sim 0.4$ with modest accuracy.  Although the
recoil from the final plunge contributes significantly to the final
recoil, we are only able to make crude estimates of its magnitude.  We
find that the recoil can easily reach $\sim 100-200\,{\rm km/s}$, but
most likely does not exceed $\sim 500\,{\rm km/s}$.  Though much
lower than previous estimates, this recoil is large enough to have
important astrophysical consequences. These include the ejection of
black holes from globular clusters, dwarf galaxies, and high-redshift
dark matter halos.
\end{abstract}


\section{Introduction and Background}

Along with energy and angular momentum, gravitational waves (GWs)
carry {\it linear} momentum away from a radiating source
{\citep{br61,peres62,bek73}}. Global conservation of momentum requires that
the center of mass (COM) of the system recoil. This recoil is
independent of the system's total mass.


{\citet{f83}} first computed GW recoil for binaries.  He treated the
members as non-spinning point masses ($m_1,m_2$), the gravitational
force as Newtonian, and included only the lowest GW multipoles needed
for momentum ejection.  For circular orbits Fitchett's recoil is
\begin{equation}
V_F \simeq 1480\,\mbox{km/s}\,{f(q)\over f_{\rm max}}\left({2 G M / c^2
\over r_{\rm term}}\right)^4\;,
\label{eq:vfitchett}
\end{equation}
where $r_{\rm term}$ is the orbital separation where GW emission
terminates, $q = m_1/m_2 \le 1$ is the mass ratio, and $M = m_1 + m_2$ is
the total mass.  The function $f(q)=q^2(1-q)/(1+q)^5$ has a
maximum $f_{\rm max}$ at $q \simeq 0.38$, is zero
for $q = 1$, and has the limit $f(q) \approx q^2$ for $q \ll 1$.

Equation (\ref{eq:vfitchett}) tells us that in the coalescence of
binary black holes (BHs)---where $r_{\rm term}$ can approach
$GM/c^2$---the kick might reach thousands of km/s.  This is far
greater than the escape velocity of many globular clusters (typically
$\sim 30$ km/s), and may even exceed galactic escape velocities ($\sim
1000$ km/s).  Recoil could thus have important astrophysical
implications {\citep{rr89}} [some of which are explored in a companion
paper (\citealt{paperII}; Paper II)]. This has motivated us to revisit
this problem.


Equation (\ref{eq:vfitchett}) indicates that the recoil is strongest
at small separations, when the relativistic effects neglected by
Fitchett are most important. This issue has been addressed in restricted
circumstances using perturbation theory {\citep{nh83,fd84,nok87}},
post-Newtonian expansions {\citep{agw92,kidder}}, and numerical
relativity {\citep{ap97,ABprl,BAprd,lp04}}.
Unlike previous analyses, our treatment applies
to the strong-gravity, fast-motion regime around spinning holes
undergoing binary coalescence. Using BH perturbation theory
we model the dynamics of the binary, the
generation of GWs, and the backreaction of those waves on the
system up to the inner-most stable circular orbit (ISCO). Our results
are accurate only for extreme mass ratio inspirals ($q\ll 1$), but we
can extrapolate to $q \sim 0.4$ with modest error. We model the GW
emission from the final plunge more crudely.

\section{Overview of gravitational radiation recoil}
\label{sec:overview}

The rate at which momentum is radiated is given by
\begin{equation}
{dP_{\rm GW}^k\over dt} = {r^2\over 16\pi}\int d\Omega \left\langle
\dot{h}_{+}^2 + \dot{h}_{\times}^2  \right\rangle\,n^k\;,
\label{eq:pdot_general}
\end{equation}
where $h_{+,\times}$ are the ``plus'' and ``cross'' GW polarizations,
$n^k$ is a unit radial vector from the source,
and $r$ is the distance to the observer {\citep{thorne80}}.
[We have set $G=c=1$; an
overdot refers to a derivative with respect to coordinate time $t$;
angle brackets denote averaging over several wavelengths.]
The binary's COM recoil is $dP_{\rm COM}^k/dt = -dP_{\rm GW}^k/dt$.

Decomposing $h_{+,\times}$ into multipoles in the wave zone
{\citep{thorne80}}, Eq.\ (\ref{eq:pdot_general}) can be expanded (to
lowest order) as
\begin{equation}
{dP_{\rm GW}^k\over dt} = {2\over63}\left\langle {d^4 {\mathcal
I}^{ijk} \over dt^4} {d^3 {\mathcal I}^{ij}\over dt^3}\right\rangle +
{16\over45}\left\langle \epsilon^{kpq}{d^3 {\mathcal I}^{pj} \over
dt^3}{d^3 {\mathcal S}^{qj}\over dt^3}\right\rangle\;,
\label{eq:recoil_leading}
\end{equation}
where ${\mathcal I}^{ij}$, ${\mathcal S}^{ij}$, and ${\mathcal
I}^{ijk}$ are the symmetric, trace-free mass quadrupole, current
quadrupole, and mass octupole moments. Recoil thus arises
from ``beating'' between different multipoles. Applying
Eq.\ (\ref{eq:recoil_leading}) to a Newtonian binary and
integrating yields Eq.\ (\ref{eq:vfitchett}).

\begin{figure}
\psfig{figure=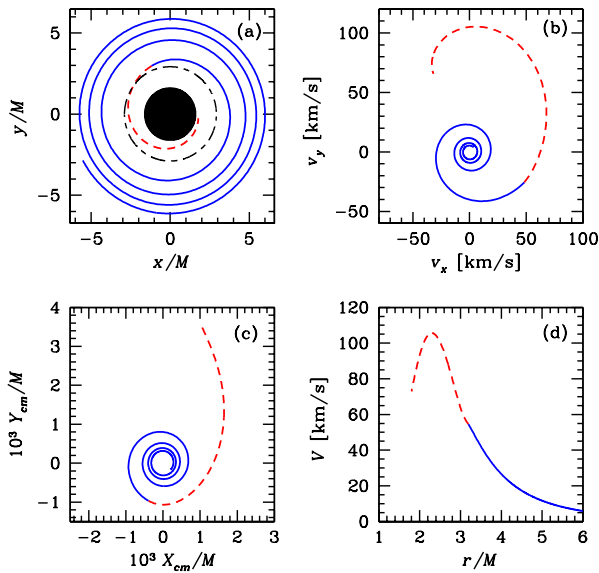,width=3.2in}
\caption{Recoil from prograde coalescence with $a/M=0.8$, $\eta=0.1$
($q=0.127$).  Solid (blue) lines represent quantities during the
inspiral, as calculated using our Teukolsky equation solver.  Dashed
(red) lines are calculations during the plunge (using the
``upper-limit'' prescription discussed in section \ref{sec:plunge}).
The plunge is
truncated shortly before the particle enters the event horizon. The
different panels are: (a) Orbit of the mass $\mu$ about the central
spinning hole. The dashed circle is the location of the ISCO. (b)
Recoil velocity of the center of mass.  The spiral ends when GW
emission is cut off. (c) Motion of the binary's center of mass.
(d) Total center of mass recoil velocity, $({v_x^2+v_y^2})^{1/2}$. }
\label{fig:a0.8pro}
\end{figure}

{\citet{agw92}} provides an intuitive description of the recoil: When
two non-spinning bodies are in circular orbit, the lighter mass moves
faster and is more effective at ``forward beaming'' its radiation.
Net momentum is ejected in the direction of the lighter mass's
velocity, with opposing COM recoil.  When $m_1 = m_2$, the beaming is
symmetric and the recoil vanishes.  The instantaneous recoil
continually changes direction over a circular orbit, so the COM traces
a circle.  Neglecting radiation reaction, this circle closes, and the
recoil averages to zero over each orbit.  With radiative losses, the
orbit does not close, and the recoil accumulates.  This accumulation
proceeds until the holes plunge and merge, shutting off the radiated
momentum flux and yielding a net, non-zero kick velocity (cf.~Fig.\
\ref{fig:a0.8pro}).  \placefigure{fig:a0.8pro}

Spin complicates this picture by breaking the binary's symmetry.
Consider an equal-mass binary, with one member spinning parallel to
the orbital angular momentum.  Due to spin-induced frame dragging, the
non-spinning body's speed---and hence radiation beaming---is enhanced.
{\citet{kidder}} has treated this spin-orbit interaction in
post-Newtonian theory.  Specializing his Eq.\ (3.31) to a circular,
non-precessing orbit, the total kick for two bodies with spins
${\mathbf S}_{1,2} = \tilde{a}_{1,2}m_{1,2}^2 \hat{{\mathbf z}}$
parallel (or antiparallel) to the orbital angular momentum is
\begin{equation}
V_{\rm kick} = \left| V_F + 883\,{\rm km/s}\,
\frac{f_{\rm SO}(q,{\tilde a}_1,{\tilde a}_2)}{f_{\rm SO,\,max}}
\left( \frac{2M}{r_{\rm term}} \right)^{9/2} \right| \;,
\label{eq:SOkick}
\end{equation}
where the spin-orbit scaling function $f_{\rm SO}(q,{\tilde
a}_1,{\tilde a_2}) = q^2({\tilde a}_2 - q{\tilde a}_1)/(1 + q)^5$.
The ``correction'' causes significant recoil even when $q = 1$ (and
hence $V_F = 0$).  The spin-orbit term is largest when $q=1$ and
the spins are maximal and antiparallel (${\tilde a_1} = -{\tilde a}_2
= \pm 1; f_{\rm SO,\,max}\equiv 1/16$).  The recoil vanishes for $q
= 1$ and spins equal and parallel (${\tilde a}_1 = {\tilde a_2}$)---a
symmetric binary.

Since we work in the $q \ll 1$ limit, we ignore the smaller body's
spin, which incurs an error $\sim q^2 \tilde{a}_1$ in the orbital
dynamics {\citep{kidder}}.  Our extreme mass ratio analysis treats
the binary in an
effective-one-body sense: a non-spinning point particle with mass
$\mu=m_1 m_2/M$ orbits a Kerr hole with mass $M=m_1+ m_2$ and spin
${\mathbf S}=\tilde{a} M^2 \hat{{\mathbf z}}$. There is an ambiguity,
however, in how one translates the physical spin parameter ${\tilde
a}_2$ of the hole to the ``effective'' spin parameter ${\tilde
a}$. {\citet{damour}} provides a relation between these parameters,
valid in the post-Newtonian limit for ${\tilde a}<0.3$: ${\tilde a} =
{\tilde a}_2 (1 + 3q/4)/(1 + q)^2$. Because of this ambiguity, we
present our results in terms of the effective-spin-parameter ${\tilde
a}$. Even if the larger hole's spin is nearly maximal (${\tilde a}_2
\simeq \pm 1$), finite mass ratios $q \gtrsim 0.1$ restrict our results to
spins with $|{\tilde a}| \lesssim 0.8-0.9$.

When applied to a perturbation calculation of the head-on collision of
two BHs, an effective-one-body scaling of the GW energy flux
($\dot{E}_{\rm GW} \propto q^2$) in which $q \rightarrow \eta =
\mu/M=q/(1+q)^2$ has been shown to agree with results from full
numerical relativity {\citep{smarr}}.
We use a similar ``scaling up''
procedure for the momentum flux: In perturbation theory
$\dot{P}^j_{\rm GW} \propto q^2$.  We then substitute $q^2\rightarrow
f(q)$ {\citep{fd84}}.  [In terms of $\eta$, the scaling function is
given by $f(q)\to f(\eta)=\eta^2 \sqrt{1-4\eta}$, and is maximized at
$\eta=1/5$.]  Using $f(q)$ [or $f(\eta)$] to scale the momentum flux
assumes both
bodies are non-spinning and that the orbit is quasi-circular.
For simplicity, approximate spin corrections to $f(q)$ based on
Eq.\ (\ref{eq:SOkick}) are ignored (incurring errors
$\lesssim 30\%$ if $q\lesssim 0.4$) (cf.~Paper II).

\section{Inspiral recoil from perturbation theory}
\label{sec:perturb}

Our model binary consists of a mass $\mu$ in circular, equatorial
orbit about a BH with mass $M$ and effective spin $a = {\tilde
a}M$.  (GWs rapidly reduce eccentricity, so circularity is a good
assumption for many astrophysical binaries.)  When $\mu \ll M$,
binary evolution is well described using BH perturbation theory
{\citep{teuk73}}.  We treat the binary's spacetime as a Kerr BH plus
corrections from solving the perturbed Einstein equations---the
Teukolsky equation.  Specifically, we solve a linear wave equation for
the complex scalar function $\Psi_4$, which describes radiative
perturbations to the hole's curvature.  Far from the binary, $\Psi_4 =
(\ddot{h}_+ - i\ddot{h}_\times)/2$; it therefore encodes information
about the GW fields in the distant wave zone, as well as the energy,
momentum, and angular momentum carried by those fields.

Far from the binary $\Psi_4$ has the expansion:
\begin{equation}
\Psi_4 = {1\over r}\sum_{lm} Z_{lm} S_{lm}(\theta;a\omega_{m})
e^{im\phi - i\omega_mt_R}\;.
\label{eq:multipole}
\end{equation}
In terms of Boyer-Lindquist coordinates $(t,r,\theta,\phi)$, $t_R=t-r$
is retarded time, $\omega_m = m\Omega_{\rm orb}$ is a harmonic of the
orbital frequency, $S_{lm}(\theta;a\omega_m)$ is a spheroidal
harmonic, and $Z_{lm}$ is a complex number found by solving a
particular ordinary differential equation \citep{circI}.

The linear momentum flux can be extracted by combining Eqs.\
(\ref{eq:pdot_general}) and (\ref{eq:multipole}). The resulting
expression is simplest in the ``corotating'' frame, $\phi^{\rm
corot}=\phi(t)-\Omega_{\rm orb} t$:
\begin{equation}
\dot{\mathcal P}_{\rm GW} = \frac{1}{2} \sum_{ll'm}
\frac{Z_{lm}{\bar Z}_{l' (m+1)}}{\omega_m \omega_{m+1}} \int_0^{\pi}
S_{lm} S_{l'(m+1)} \sin^2\theta d\theta \;.
\label{eq:pplusminus}
\end{equation}
Here, $\dot{\mathcal P}_{\rm GW}=e^{-i\phi(t)} [\dot{P}^x_{\rm GW} + i
\dot{P}^y_{\rm GW}]$, and an overbar denotes complex
conjugation. Similar expressions give the energy and angular momentum
fluxes. The recoil velocity is found by integrating Eq.\
(\ref{eq:pplusminus}), starting at initial time $T_0$ when the binary
is at large separation [and the recoil is well described by Eq.\
(\ref{eq:vfitchett})], and ending at time $T$ when GW emission
terminates:
\begin{equation}
v_x + i v_y = -{1\over M}\int_{T_0}^T e^{i\phi(t)} \dot{\mathcal
P}_{\rm GW} \,dt \;.
\label{eq:recoil_vel}
\end{equation}

Our procedure starts with a point-source on a circular geodesic orbit
with specified energy $E$ and angular momentum $L_z$. Solving the
Teukolsky equation gives us the energy, momentum, and angular momentum
fluxes of GWs to infinity and down the event horizon. (The linear
momentum flux down the horizon does not affect the recoil.) In the
adiabatic limit (in which GW backreaction changes the orbit very
slowly, $r/\dot{r} \ll 2\pi/\Omega_{\rm orb}$), the energy and angular
momentum fluxes ($\dot{E}_{\rm GW}, \dot{L}_{z,{\rm GW}}$) are used to
evolve to a new geodesic with $E-\dot{E}_{\rm GW} \Delta t$ and
$L_z-\dot{L}_{z,{\rm GW}} \Delta t$. Repeating this procedure for a
sequence of geodesics generates a slow inspiral trajectory. The
momentum flux along this trajectory and associated recoil velocity are
then calculated via Eqs.\ (\ref{eq:pplusminus}) and
(\ref{eq:recoil_vel}).

This prescription can be used to calculate the recoil velocity only up
to the ISCO. There the slow, adiabatic inspiral of the particle
transitions to a rapid ``plunge'' that terminates when the particle
crosses the event horizon
(cf.\ Fig.\ \ref{fig:a0.8pro}a). Our Fourier decomposition of $\Psi_4$
is no longer valid as there are no well-defined harmonics $\omega_m$
for plunging trajectories.

\begin{figure}
\psfig{figure=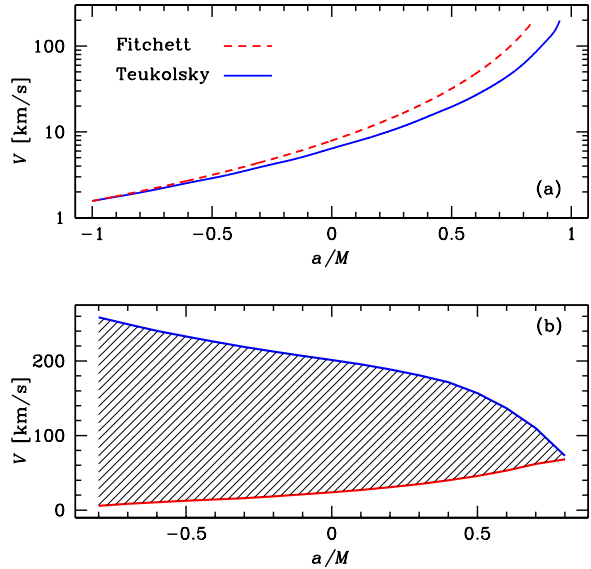,width=3.2in}
\caption{Recoil velocity versus effective spin $a/M$ for $\eta=0.1$
($q=0.127$).  (a) Recoil velocity up to the ISCO.  The solid (blue)
curve is our Teukolsky equation result.  The dashed (red) curve shows
the Newtonian recoil prediction [Eq.\ (\ref{eq:vfitchett})], which is
substantially higher for large, prograde spins (smaller ISCO radius).
(b) Upper and lower limits for the total recoil.  The shaded region
represents our uncertainty in the final kick velocity.
The detailed shape of the upper-limit curve depends on the nature of
our truncated-power-law ansatz.}
\label{fig:vrecoil}
\end{figure}

\placefigure{fig:vrecoil}

Figure \ref{fig:vrecoil}a shows the perturbation theory
calculation of the ISCO recoil for a binary with reduced mass ratio
$\eta=0.1$ ($q=0.127$). The solid curve in Figure 1a can be fit by
\begin{equation}
V_{\rm isco}=422\, {\rm km/s} \, \frac{f(q)}{f_{\rm max}} \left(
\frac{2M}{r_{\rm isco}} \right)^{2.63+0.06 r_{\rm isco}/M} \,,
\label{eq:fitvisco}
\end{equation}
where $r_{\rm isco}$ is the spin-dependent ISCO radius [defined for
$q=0$ by Eq.\ 2.21 of {\citet{bpt}}], and we have included the
appropriate scaling function
(valid for $q\lesssim 0.4$).
Although our adiabatic assumption is violated for $\eta=0.1$
(especially for large, prograde spins) our results are still valid
since $V_{\rm isco}/f(q)$ is only weakly dependent on $q$ (and is
independent of $q$ in the $q \rightarrow 0$ limit).

For retrograde orbits around rapidly spinning holes, the ISCO is at
large radius ($9M$ for ${\tilde a}=-1$) and Fitchett's Newtonian
formula [Eq.\ (\ref{eq:vfitchett})] agrees well with our
result. For prograde inspiral into
rapidly spinning holes, the ISCO is deep in the strong
field, where relativistic effects become important and suppress the
recoil relative to Fitchett's result.

\section{Recoil estimates from the final plunge}
\label{sec:plunge}

During the plunge, the small body's motion is dominated by the Kerr
effective potential rather than radiation-reaction forces
{\citep{ot00}}.  It is easy
to match a plunging geodesic with constant $E$ and $L_z$ onto an
inspiral trajectory near the ISCO.  With a code that does not Fourier
expand $\Psi_4$ {\citep{khanna,martel}}, one could properly compute
the GW emission and associated recoil along such a plunging trajectory
(when $q \ll 1$).

Since we do not have such a code at hand, we must estimate the wave
emission more crudely. Our results from the inspiral show that, for a
given spin, $\dot{\mathcal P}_{\rm GW}$ is well described by a power
law in radius, $\dot{\mathcal P}_{\rm GW}\propto r^{-\alpha}$, from
large $r$ up to the ISCO.  As an approximate ``upper limit'' of the
recoil, we make the ansatz that this power law can be continued past
the ISCO. This must break down at some point: the power-law reflects
the circularity of the inspiral orbit and should be suppressed by the
increasingly radial motion during the plunge. To prevent the momentum
flux from diverging, we truncate the power law at $3M$, replacing it
with the condition that $(dt/d\tau)\dot{\mathcal P}_{\rm GW} = {\rm
constant}$, where $\tau$ is proper time along the plunge
geodesic. This allows the momentum flux to ``redshift away'' as the
particle approaches the horizon. Using the recoil velocity at the ISCO
as initial conditions [Sec.\ \ref{sec:perturb}] and a plunge
trajectory with coordinates $[r(t),\phi(t)]$, we use Eq.\
(\ref{eq:recoil_vel}) and our truncated-power-law ansatz to compute
the accumulated recoil until a cutoff time $T$ when the horizons of
the holes come into contact (in a quasi-Newtonian interpretation of
the coordinates). The upper-curve of Figure \ref{fig:vrecoil}b shows
the result of this calculation (for $\eta=0.1$).

We also perform a separate ``lower-limit'' calculation. A plunge
trajectory is computed as before, but in place of the power-law ansatz
for $\dot{\mathcal P}_{\rm GW}$, we integrate the truncated, multipole
expansion of Eq.\ (\ref{eq:recoil_leading}) instead. In this
calculation the momentum flux initially grows like a power law, but
then decreases as the plunging trajectory nears the event
horizon. Because we neglect higher multipoles (which are extremely
important in the fast-motion, strong-gravity region), this method
likely underestimates the recoil. The total accumulated recoil at the
cutoff time $T$ using this method is shown in the lower curve of
Figure \ref{fig:vrecoil}b (also for $\eta=0.1$).

The shaded region between the two curves in Figure \ref{fig:vrecoil}b
represents our uncertainty in the total recoil at the end of the
plunge. This uncertainty is largest for retrograde orbits around
rapidly spinning holes, in which the distance the particle must
``plunge'' is greatest. For prograde inspiral into rapidly spinning
holes, much of the recoil is due to emission during the slow inspiral
phase, for which our BH perturbation techniques are well-suited.
Figure \ref{fig:a0.8pro} shows the relative contributions from the
inspiral and plunge for such a scenario.

Although the two calculations for the plunge recoil give rather
different results, useful astrophysical information is contained in
the approximate upper and lower bounds that they represent. The
estimate $V \sim 120\,{\rm km/s}$ bisects the shaded region of Figure
\ref{fig:vrecoil}b and represents a typical recoil velocity for this
mass ratio. Note also that the numbers in Figure \ref{fig:vrecoil} can
be scaled to higher mass ratios by multiplying by $f(q)/f(\eta =
0.1)$. For $q\approx 0.38$ this implies that our results can be
augmented by a factor $\approx 2.3$.

\section{Discussion}

The punchline of this analysis is simple: quasi-Newtonian estimates have
significantly overestimated the kick velocity from anisotropic GW
emission during binary coalescence.  The recoil is strongest when the
smaller member is
deep in the strong-field of the large black hole.  General relativistic
effects, such as the gravitational redshift and spacetime
curvature-scattering, act on the emitted GWs and reduce the recoil.

Though reduced, the recoil remains large enough to have important
astrophysical consequences. Recoils with $V \sim 10$--$100\, {\rm
km/s}$ are likely; kicks of a few hundred km/s are not
unexpected; and the largest possible recoils are probably
$\lesssim 500\,{\rm km/s}$.  These speeds are smaller than most
galactic escape velocities, suggesting that
BH mergers that follow galaxy mergers will remain within their host
structures.  However, these recoils are similar to the escape
speeds of dwarf galaxies; and they may be sufficient to escape
from mergers in high redshift structures [$z\gtrsim 5-10$; cf.\
{\citet{bl01}}, Fig.\ 8].
Binary BH ejection from globular clusters is quite likely, with
significant implications for the formation of intermediate mass
black holes (IMBH) via hierarchical mergers {\citep{mc03}}.
Our recoil estimates will also be useful in
simulations of supermassive and IMBH evolution in dark halos
{\citep{volonteri1,volonteri2}}.

Future papers will present the formalism used for this analysis, and
will investigate the influence of orbital inclination on the
recoil. More work in perturbation theory also remains in addressing
the recoil from the plunge and final ringdown of the merging black
holes.

Finally, \citet{rr89} have speculated that spin-orbit misalignment
could lead to recoil directed out of the orbital plane. This recoil
might accumulate secularly rather than oscillate, and would be similar
to the ``electromagnetic rocket'' in pulsars with
off-centered magnetic dipole moments {\citep{ht75,lcc01}}.  We suspect
that this effect occurs but it is likely small compared to the
recoil from the final plunge and merger. Firm estimates of the final
kick velocity will rely on correctly modelling the final phase of BH
coalescence.  For comparable mass binaries, full numerical relativity
will ultimately be needed to accurately compute the GW recoil.

\acknowledgements

We thank Saul Teukolsky and Jerry Ostriker for bringing this problem
to our attention.  For helpful discussions, we thank Avi Loeb, David
Merritt, Milo\v{s} Milosavljevi\'{c}, Martin Rees, Joseph Silk, Alan
Wiseman, Yanqin Wu, and most especially, \'Eanna Flanagan. We
gratefully acknowledge the support of the Kavli Institute for
Theoretical Physics, where this work was initiated. MF is supported by
NSF Grant PHY-0140209; SAH by NASA Grant NAG5-12906 and NSF Grant
PHY-0244424; and DEH by NSF Grant PHY-0114422.

\end{document}